\documentclass[aps,prd,twocolumn,twoside,superscriptaddress,floatfix, nofootinbib]{revtex4-1}
\pdfoutput=1
\usepackage{ifthen,amsmath,amssymb, bm}
\usepackage{graphicx}
\usepackage{color}
\usepackage{hyperref}
\usepackage{float}
\usepackage{aas_macros}

\newcommand{\vk} {\bm{k}}

\newcommand{\vn} {\hat{\bm{n}}}

\newcommand{\beq} {\begin{equation}}
\newcommand{\eeq} {\end{equation}}
\newcommand{\bal} {\begin{aligned}}
\newcommand{\eal} {\end{aligned}}

\begin{document}

\title{Is the Rees-Sciama effect detectable by the next generation of cosmological experiments?}

\author{Simone Ferraro}
\email{sferraro@lbl.gov}
\affiliation{Lawrence Berkeley National Laboratory, One Cyclotron Road, Berkeley, CA 94720, USA}
\affiliation{Berkeley Center for Cosmological Physics, Department of Physics,
University of California, Berkeley, CA 94720, USA}
\author{Emmanuel Schaan}
\affiliation{Lawrence Berkeley National Laboratory, One Cyclotron Road, Berkeley, CA 94720, USA}
\affiliation{Berkeley Center for Cosmological Physics, Department of Physics,
University of California, Berkeley, CA 94720, USA}
\author{Elena Pierpaoli}
\affiliation{Physics \& Astronomy Department, University of Southern California, Los Angeles, California, 90089-0484}

\begin{abstract}
Non-linear growth of structure causes the gravitational potentials to grow with time, and this leaves an imprint on the small-scale temperature fluctuations of the Cosmic Microwave Background (CMB), a signal known as the Rees-Sciama (RS) effect. Building on previous ``idealized'' studies, here we investigate the detectability of the RS effect by cross-correlating upcoming CMB and Large-Scale Structure surveys. We include tracers with realistic number density and bias, realistic noise for upcoming and future CMB experiments, and importantly, the contribution from CMB foregrounds to the noise budget. We also derive optimal redshift weights, which are crucial to the detection due to the mismatch between the redshift kernel of the RS effect and the typical redshift distribution of current and upcoming galaxy surveys. In agreement with previous work, we confirm that the signal would in principle be detectable at high significance by ``white noise'' versions of future CMB experiments, when foregrounds are not included as part of the noise. However, we show that inclusion of foregrounds limits the statistical detectability of the signal: an optimally-weighted high-redshift sample from Rubin Observatory LSST, together with CMB maps from CMB-S4 or CMB-HD, can yield a detection with signal-to-noise 6 - 8, when taking $\ell_{\rm max} = 6000$, provided that foreground-induced biases can be successfully controlled. Improvements are possible if the total power from foregrounds is further reduced, for example by more aggressive masking, or if the signal can be modelled down to smaller scales.

\end{abstract}

\maketitle

\section{Introduction}
The Integrate Sachs-Wolfe (ISW) effect \cite{1967ApJ...147...73S} is one of the gravitational contributions to the observed Cosmic Microwave Background (CMB) Temperature fluctuations.  It arises from the time-dependence of the gravitational potential as photons propagate from the surface of last scattering to us.
Depending on the regime of interest and the physical origin of the time-evolution of the potential, the same mechanism for producing ISW fluctuations is sometimes known as the Rees-Sciama (RS) effect \cite{1968Natur.217..511R}. While ISW and RS have often been used interchangeably, it appears that in recent literature ISW has been predominantly used to refer to the large-scale, linear effect, where the time evolution of the potential is due to the effect of Dark Energy \cite{Afshordi:2004kz, Nishizawa:2014vga, Ferraro:2014msa, Krolewski:2021znk, Planck:2015fcm, Giannantonio:2008zi, Giannantonio:2012aa}. At the same time, RS has usually referred to the non-linear contributions, for example from the non-linear collapse of matter around galaxies and clusters, which causes the (absolute value of) the potential to grow \cite{Nishizawa:2014vga, Seljak:1995eu, Smith:2009pn}.  We'll adopt the convention above in this work, noting that there is no universal consensus on the terminology. 

We also point out that another source of time-varying potential is the galaxy motion across the line of sight \cite{1983Natur.302..315B}, generating a dipole signal in the CMB, aligned with the direction of motion, an effect known as ``moving lens'' or ``slingshot''. This component of the signal and its detectability has been extensively studied in recent years \cite{Stebbins:2006tv, Hotinli:2018yyc, Hotinli:2020ntd, Hotinli:2021hih, Yasini:2018rrl, Hagala:2019ncx}, and that due to the dipole nature of the signal, it doesn't contribute to the cross-correlations studied in this paper. Therefore the moving lens effect won't be explored here. 

In this work we will only consider the effect from non-linear growth of structure, and we'll only include this contribution when referring to RS: by directly probing the growth of the gravitational potential on small scales, it is potentially an excellent probe of gravity in the non-linear regime, often sensitive to modifications to General Relativity \cite{Nishizawa:2014vga}.

The RS signal in a Cold Dark Matter (CDM) Universe was studied in detail  in \cite{Seljak:1995eu}, with the help of second-order perturbation theory and N-body simulations. The author finds that the RS contribution is negligible in the CMB power spectrum on large scales, but that is a potentially significant source of anisotropy on subarcminute scales ($\ell \gtrsim 5000$). The effects of non-linearities in matter and bias, as well as its redshift dependence was further studied in \cite{Smith:2009pn}. The authors further point out that the RS signal is suppressed at late times during Dark Energy domination due to the suppression in growth caused by the accelerated expansion. Indeed, this will be evident in Section \ref{sec:where_signal}, where we show that most of the signal arises during matter domination at $z \gtrsim 1$. More recent simulations of the RS effect are also found in \cite{Cai:2010hx, Carbone:2016nzj}, which do not forecast the detectability of the signal.

The cross-correlation between the RS effect and weak lensing has been explored in \cite{Nishizawa:2007pv, Xu:2016jns, Lee:2015rna}, showing promise for moderate-to-high significance detections with the next generation of cosmological surveys, under idealized assumptions about the CMB noise, which usually neglect the effect of the foregrounds. While we won't specifically forecast the use of weak lensing as a Large Scale Structure (LSS) tracer, our ``white noise'' forecasts in Section \ref{sec:warmup} are in qualitative agreement with the references just mentioned.
Finally, a comprehensive review of the ISW and RS effects, as well as perturbative methods to evaluate the RS contribution can be found in \cite{Nishizawa:2014vga}.

In this paper, we use our modern understanding of the small-scale fluctuations of the CMB and of high-redshift galaxies to provide realistic forecasts for the next generation of experiments. Our paper differs from previous work in the following three ways: (1) We include the contributions of foregrounds and the channel-by-channel noise properties of upcoming experiments, optimally combined to reduce the total background fluctuations, to go beyond the idealized ``white noise'' projections of previous work: this will be shown to lead to major changes in the forecasts. (2) We also explore realistic LSS tracers in the form of high redshift samples derived from the upcoming Rubin Observatory LSST, which includes dropout samples characterized from recent observations and (3) we derive and use optimal redshift weighting for the LSS tracers, which will be crucial to the detectability of the signal. Throughout the paper, we shall use Planck 2018 cosmological parameters, from Table 2 (last column) of \cite{Planck:2018vyg}.

The remainder of the paper is organized as follows: in Section \ref{sec:RS} we review the calculation of the RS effect. In Section \ref{sec:warmup} we consider the case of an ``idealized'' cosmic-variance limited LSS survey and a white noise CMB experiment to connect to previous literature, while in Section \ref{sec:where_signal} we explore the redshift origin of the signal. In Section \ref{sec:ILC} we discuss the CMB experiments considered here (CMB-S4 and CMB-HD), while Section \ref{sec:LSS} we describe our fiducial tracers of LSS. Finally we show our results in Section \ref{sec:results} and conclude in Section \ref{sec:conclusions}. Appendix \ref{app:weights} derives the optimal weights used in the text.

\section{The Rees-Sciama effect}
\label{sec:RS}
When photons propagate in a time-varying gravitational potential $\Phi$, they undergo a change in energy, which in units of CMB temperature is given by
\begin{equation}
    \left( \frac{\Delta T}{T} \right)_{\rm ISW}(\vn) = -2 \int d \chi e^{-\tau(z)} \dot{\Phi}(\vn \chi)
    \label{eq:ISW}
\end{equation}
Here $\tau(z)$ is the mean optical depth to redshift $z$,
$\chi(z)$ is the comoving distance, the dot denotes the derivative with respect to conformal time, and we have adopted natural units where $c=1$.
The gravitational potential is in turn related to the matter fluctuations $\delta_m$ through the Poisson equation: 
\begin{equation}
    k^2 \Phi = \frac32 \Omega_{m,0} H_0^2 \frac{\delta_m}{a}
\end{equation}
where $a = 1/(1+z)$ is the scale factor, $H_0$ is the Hubble parameter and $\Omega_{m,0}$ is the matter fraction, both evaluated at $z=0$. 
In the linear regime, during matter domination, it is well known that $\delta_m \propto a$, so that $\Phi$ is constant and the linear part of ISW vanishes. 
Note that the presence of Dark Energy is not required to get a RS effect: non-linear growth which differs from $\delta_m \propto a$ will lead to a non-zero RS signal. Dark Energy causes $\delta_m$ to grow slower than $a$ on large scales at late times, generating a positive correlation between tracers such as galaxies and CMB fluctuations (the ISW effect). On sufficiently small non-linear scales, gravitational collapse causes $\delta_m$ to grow faster than $a$, creating a negative correlation between tracers and the CMB (the RS effect). The presence of Dark Energy is not required for this latter part of the signal to be non-zero, and in fact we shall see that the majority of the signal originates from the matter-dominated era.

We can rewrite Poisson's equation as:
\begin{equation}
    \delta_m(\vk, z) = \alpha^{\rm ISW}(k,z) \Phi(\vk,z)
\end{equation}
with
\begin{equation}
    \alpha^{\rm ISW}(k,z) = \frac23 \frac{k^2}{\Omega_{m,0} H_0^2 (1+z)}
\end{equation}
Now let's consider density fluctuations  $\delta_0(\vk) = \delta_m(\vk, z_0)$, at some \textit{fixed} but arbitrary $z_0$. Then we can define a scale dependent growth function in terms of the non-linear and linear matter power spectra ($P^{\rm NL}_\delta$ and $P^{\rm lin}_\delta$ respectively) as:
\begin{equation}
    D(k, z) = \sqrt{\frac{P^{\rm NL}_\delta(k, z)}{P^{\rm lin}_\delta(k, z_0)}}
\end{equation}
Note that the numerical value of $D$ depends on the choice of $z_0$, but that our results will be independent of the normalization of $D$ and therefore of $z_0$. One note is in order here: outside of the linear regime, it is no longer exact that $\delta_m(\vk,z) = D(k,z) \delta_0(\vk)$ due to the decorrelation induced by gravitational non-linearities on small scales. Nevertheless, one may conjecture that this approximately holds even in the non-linear regime when considering expectation values (for example power spectra or correlation functions). Indeed, a mathematically equivalent approximation has been tested with the help of N-body simulations by \cite{Nishizawa:2007pv} and shown to work at the $\approx 10\%$ level, and therefore we shall replace $\delta_m(\vk,z) \rightarrow  D(k,z) \delta_0(\vk)$ when calculating power spectra. 

As a practical example, we shall consider galaxy overdensity as our LSS tracer of choice, while noting that the same formalism applies to any tracer of the LSS such as galaxy lensing or Intensity Mapping. The galaxy redshift distribution is defined as $n(z) \equiv dN / (dz\ d\Omega)$, where $N$ is the galaxy number counts, 
and $\Omega$ is the solid angle. Then the mean projected number density is given by $\bar{n}_{\rm 2D} = \int dz \ n(z)$ and the galaxy overdensity in direction $\vn$ is 
\begin{equation}
    \delta_g(\vn) \equiv \frac{\delta n_{\rm 2D}(\vn)}{\bar{n}_{\rm 2D}} = \int dz\ W^g(z) \delta_m(\chi \vn, z)
    \label{eq:galaxies}
\end{equation}
with 
\begin{equation}
    W^g(z) = b_g(z) \frac{n(z)}{\bar{n}_{\rm 2D}}
\end{equation}
Here $b_g(z)$ is the linear bias\footnote{On small scales, a linear bias approximation will not be adequate for precision cosmological analyses \cite{Senatore:2014eva, Modi:2017wds, Desjacques:2016bnm}. For the purpose of our forecast for detectability we shall consider linear bias only, while leaving a full analysis of bias non-linearities to future work.} of the galaxy sample used in the cross-correlation.

Using Equation \ref{eq:ISW}, together with Equation \ref{eq:galaxies}  and the Limber approximation \cite{1953ApJ...117..134L, LoVerde:2008re}, we can  calculate the cross-correlation  $C_{\ell}^{Tg}$ between $(\Delta T / T)_{\rm ISW}$ and $\delta_g$:
\begin{equation}
    C_{\ell}^{Tg} = \int dz \frac{H(z)}{\chi^2} W^{g}(z) W^{\rm ISW}(k = \ell/\chi, z) P^{\rm NL}_{\delta}(k = \ell/\chi, z)
    \label{eq:RS_Limber}
\end{equation}
where 
\begin{equation}
    W^{\rm ISW}(k, z) = \frac{2 e^{-\tau(z)}}{D(k,z)} \frac{d}{dz}\left( \frac{D(k,z)}{\alpha^{\rm ISW}(k,z)} \right) 
    \label{eq:WISW}
\end{equation}

\begin{figure}
    \centering
    \includegraphics[width=\columnwidth]{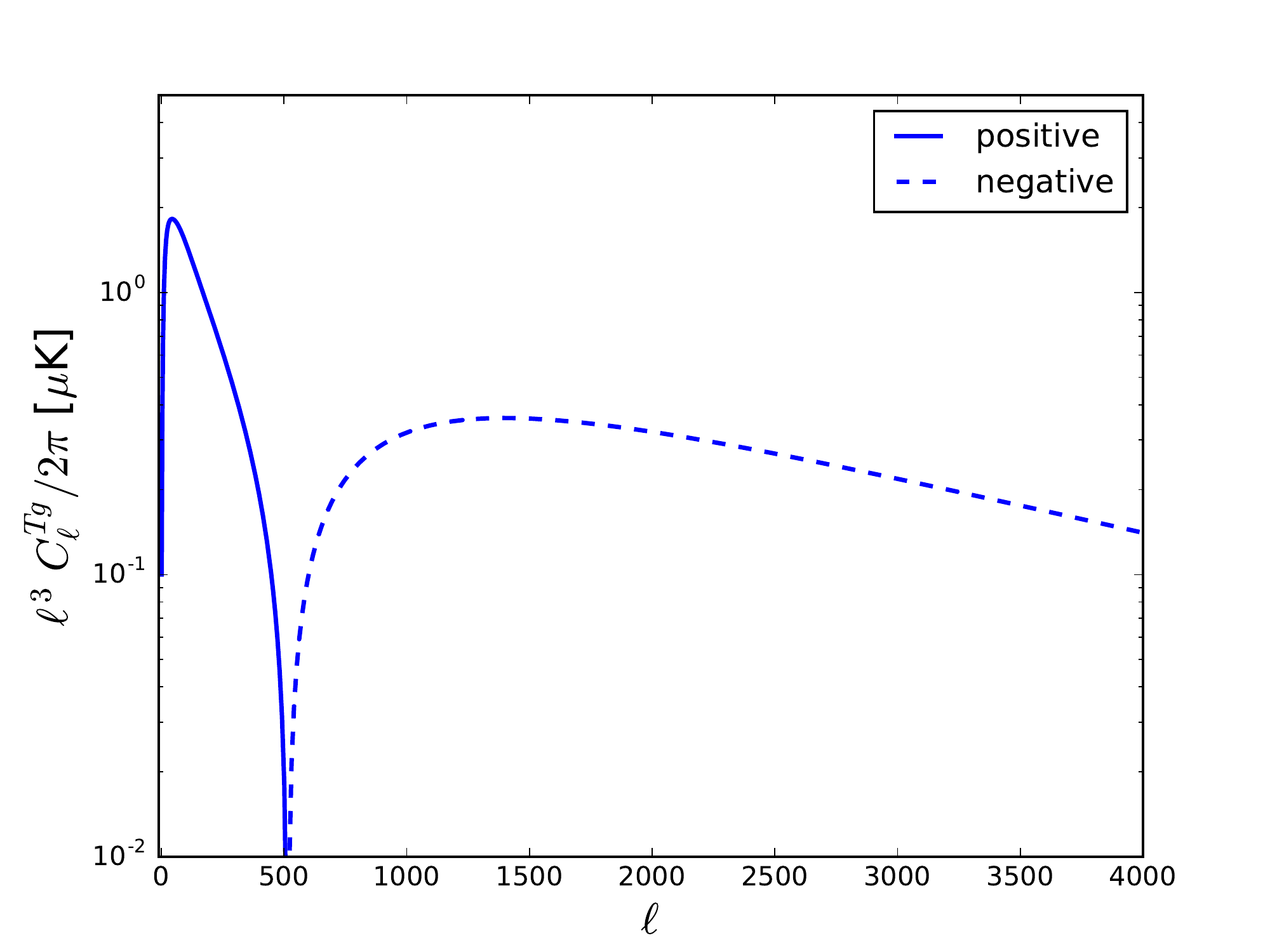}
    \caption{The ISW and RS signal when cross-correlating a CMB temperature map with galaxies from the LSST ``gold'' sample: on large scales ($\ell \lesssim 500$), the signal is positive and is primarily due to linear ISW, while on smaller scales, the signal is negative, representing the effect of non-linear growth in what we call the RS effect. 
    }
    \label{fig:RS_signal}
\end{figure}

For the non-linear matter power spectrum, we shall use the \texttt{halofit} fitting function \cite{Smith:2002dz}, as updated by \cite{Takahashi:2012em} and implemented in \texttt{CAMB}\footnote{\url{https://camb.info/}} \cite{Lewis:1999bs}.

Figure \ref{fig:RS_signal} shows the expected signal when cross-correlating CMB temperature maps with the ``gold sample'' \cite{LSSTSciBook} of galaxies from LSST, taken here as an example of LSS tracer.  In what follows, we will call Rees-Sciama the ``negative'' part of the signal at $\ell \gtrsim 500$ for which non-linear growth dominates. The large scale ($\ell \lesssim 500$) positive correlation is due to the linear ISW effect, which won't be included in the forecasts (unlike in some of the previous work): this linear part is well-studied and previously detected, and its detectability within the standard $\Lambda$CDM cosmological model is limited by cosmic variance to $S/N \lesssim 7.6 \sqrt{f_{\rm sky}}$, where $f_{\rm sky}$ is the sky fraction of overlap between the CMB and LSS experiment \cite{Giannantonio:2012aa}.

\section{Warmup: white noise, cosmic-variance limited forecasts}
\label{sec:warmup}
In this section we consider forecasts for a CMB experiment with finite beam and white noise only, with no foregrounds included. This is to build intuition and to connect with previous literature, which often made this approximation.

As a LSS tracer, we shall consider a cosmic-variance (CV) limited experiment out to $z_{\rm max} = 7$, from which we can form a perfect ISW template, 100\% correlated with the RS effect in the primary CMB map up to $z_{\rm max}$. We hence take $W^g(z) = W^{\rm ISW}(z)$ up to  $z_{\rm max}$ and zero otherwise. This is because this choice maximizes the $S/N$ and hence represents the CV limit.
This could be achieved for example by a futuristic high-density galaxy survey. Alternatively, low shot noise at high redshift can also be achieved with Line Intensity Mapping (LIM) experiments such as 21cm or others lines\footnote{Although we caution that the usual challenges in cross-correlating LIM experiments with 2D fields also apply here: most of the RS signal is contained in the very low $k_\parallel$, usually the most affected by foregrounds for LIM. Therefore either excellent foreground control or non-linear transformation of the LIM data would be required.}.

Therefore, the results of this section represent the theoretical limit of what can potentially be achieved by future experiments, and is therefore worth exploring. Realistic CMB and LSS assumptions for the next generation of experiments are presented in later sections. 

The signal-to-noise ratio for the amplitude of the RS in this case is given by:

\begin{equation}
    \left( \frac{S}{N} \right)_{\rm CV}^2(\ell_{\rm max}) \approx f_{\rm sky} \sum_{\ell_{\rm min} = 500}^{\ell_{\rm max}} \frac{(2\ell+1)(C_\ell^{\rm ISW})^2} {C_\ell^{\rm ISW}C_\ell^{TT}+(C_\ell^{\rm ISW})^2}
    \label{eq:SNR_CV}
\end{equation}
Where 
\begin{equation}
    C_{\ell}^{\rm ISW} = \int dz \frac{H(z)}{\chi^2} [W^{\rm ISW} (k = \ell/\chi, z)]^2 P^{\rm NL}_{\delta}(k = \ell/\chi, z)
    \label{eq:ISW_auto}
\end{equation} 
and the CMB power spectrum $C_\ell^{TT}$ includes both the lensed primary CMB\footnote{Note that in principle a (small) reduction of the noise can be obtained by using the non-zero correlation between $T$ and $E$-mode polarization, essentially by using $E$ to predict part of the primordial $T$ fluctuations, hence reducing the cosmic variance noise. Given that they are only weakly correlated, this is expected to decrease the noise by only a few percent, and won't be pursued here, but can always be included in a real analysis.} as well as detector noise with a white power spectrum $N^{\rm det}_\ell$, uncorrelated with all of the other components and given by
\begin{equation}
N^{\rm det}_{\ell} =  \Delta^2_T e^{\theta^2_{\rm FWHM} \ell^2/(8 \ln2)} \ ,
\label{eq:N_det}
\end{equation}
where $\Delta_T$ is the noise level of the experiment (usually quoted in $\mu$K-arcmin) and $\theta_{\rm FWHM}$ is the full-width at half-maximum (FWHM) of the beam in radians.

Starting from Section \ref{sec:ILC}, the total $C_\ell^{TT}$ will also include foregrounds, as well as the minimum variance noise from several frequency channels, but the ``white noise'' forecasts in this section only include the primary CMB and white detector noise for a single channel.

\begin{figure}
    \centering
    \includegraphics[width=\columnwidth]{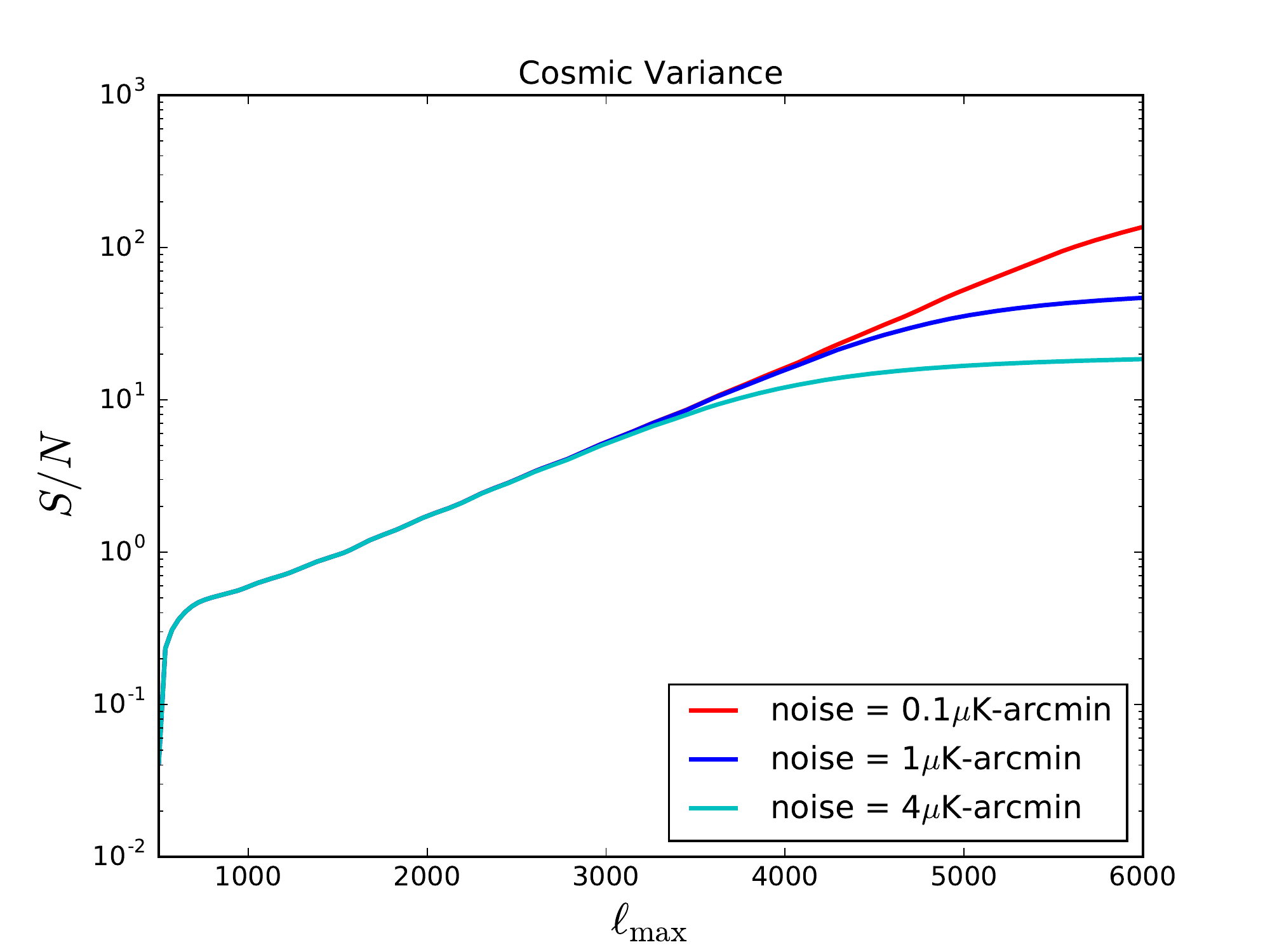}
    \caption{The cosmic variance limit for a CMB experiment with a $\theta_{\rm FWHM} = 1.4$ arcmin beam and different CMB white noise levels.}
    \label{fig:RS_signal_CV}
\end{figure}

As explained in the previous section, the lower limit to the $\ell$-range in Equation \ref{eq:SNR_CV} is taken to be $\ell_{\rm min} = 500$ in order to isolate the RS contribution. While the crossing between the linear ISW and RS is slightly dependent on the LSS sample under consideration (and in particular on its redshift distribution), for simplicity we fix $\ell_{\rm min} = 500$ irrespective of the sample. This is expected to have a minimal impact on our results since the linear ISW has negligible contributions at $\ell \gtrsim 100$, while RS receives its largest contribution from $\ell$ of a few thousand.


Figure \ref{fig:RS_signal_CV} shows the CV-limit for a low noise CMB with $\theta_{\rm FWHM} = 1.4$ arcmin.
We limit our forecasts to $\ell_{\rm max} = 6000$ due to the large uncertainties in modeling the non-linear power spectrum and its derivatives beyond that, however noting that there should definitely be a non-negligible signal at higher multipoles. Our results in Figure \ref{fig:RS_signal_CV} show that even with this conservative choice, the RS signal is, at least in principle, detectable at very high significance with future CMB and LSS experiments, and can exceed $S/N \sim 100$ for very low noise CMB experiments. We shall see that foregrounds in temperature will unfortunately limit the maximum $S/N$ obtainable, by adding a large amount of small-scale fluctuations that will act as noise in the small-scale CMB spectrum.

\section{Where does the signal come from?}
\label{sec:where_signal}
\begin{figure}[h!]
\includegraphics[width=\columnwidth]{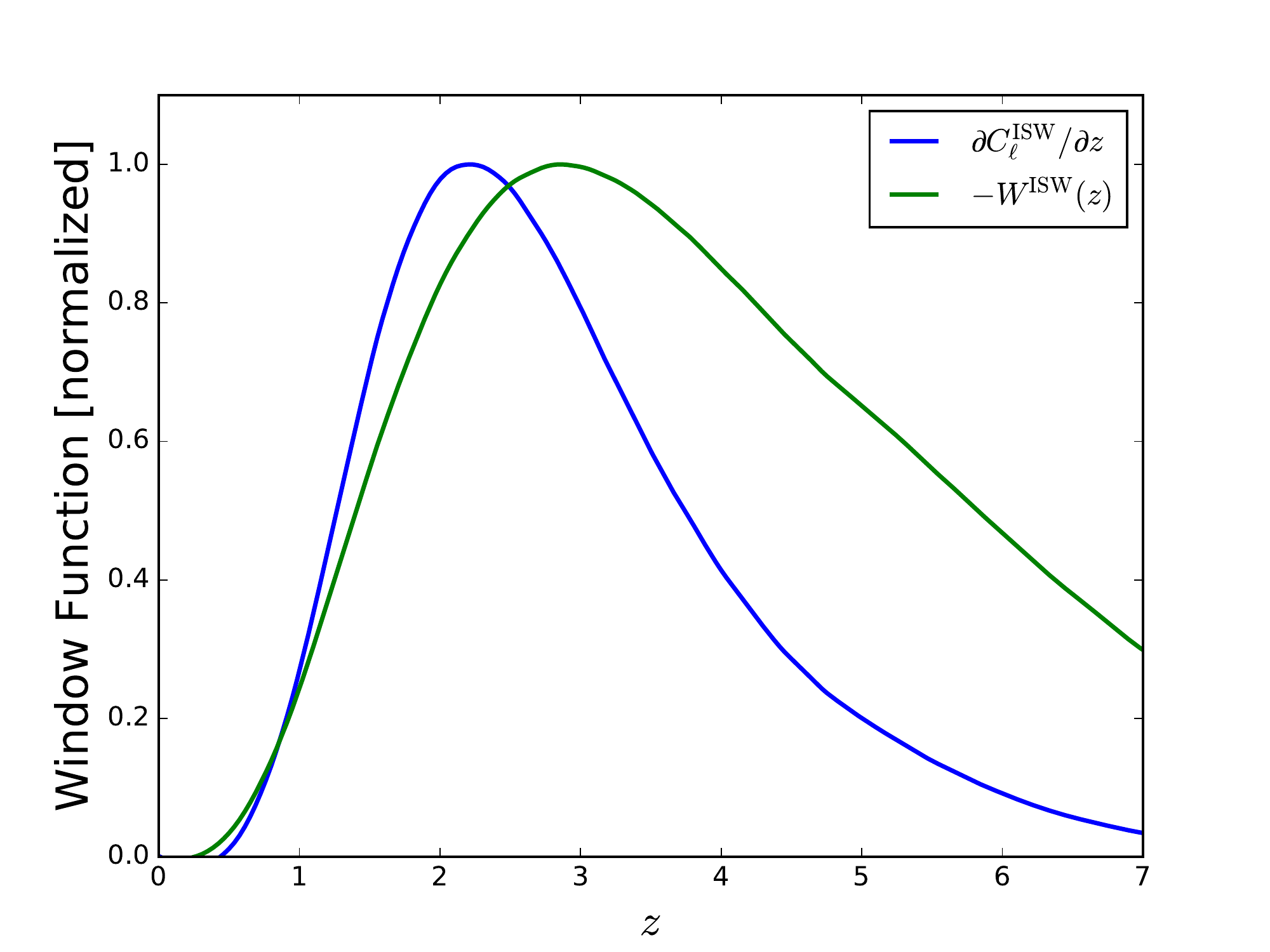}
\caption{The derivative $\partial C_\ell^{\rm ISW} /\partial z$ and the RS kernel $W^{\rm ISW}(z)$ as a function of redshift. Both are evaluated at a fixed multipole $\ell_* = 5000$.}
\label{fig:der_plot}
\end{figure}

Before discussing realistic samples and their optimal weighting, we investigate what the characteristics of an ``ideal'' tracer of RS are.
We are interested here in determining which redshifts most contribute to the signal, which is represented by $C_\ell^{\rm ISW}$.
We can see from Equation \ref{eq:ISW_auto} that $C_\ell^{\rm ISW}$ is shaped by two quantities: the kernel $W^{\rm ISW}(z)$ (Equation \ref{eq:WISW}),  and the non-linear matter power spectrum.
The kernel $W^{\rm ISW}(z)$
informs us on the 
signal induced by a unit mass fluctuation at a given redshift, while the power spectrum conveys information on  the actual amplitude of perturbations at a given redshift and scale. 
Figure \ref{fig:der_plot} shows
the kernel $W^{\rm ISW}(z)$ and the integrand $\partial C_\ell^{\rm ISW} /\partial z$ from Equation \ref{eq:ISW_auto}. Since these quantities depend on scale, and since we'll see that most of the signal will originate from multipole of a few thousands, we plot these quantities at a fixed $\ell_* = 5000$.

We can see that the kernel $W^{\rm ISW}(z)$  has very little contribution from the Dark Energy era at $z \lesssim 1$, and it peaks at $z \approx 3$, with significant contributions to rather large redshift. 

In practice, however, the size of matter fluctuations is modulated by the power spectrum and it is not uniform with redshift. This is captured when looking at $\partial C_\ell^{\rm ISW} /\partial z$: since fluctuations are larger at late times, this quantity peaks at slightly lower redshift (around $z \approx 2$) and decays faster with redshift.

The results above confirm the finding of \cite{Smith:2009pn}, which argued that the signal is heavily suppressed in the Dark Energy domination era by the competing effects of accelerated expansion and non-linear growth due to gravity. Therefore, a detection in cross-correlation with low redshift catalogs is likely to be challenging. Indeed a sample that maximizes the density of high-redshift galaxies, together with optimal weighting, will be essential for a detection as we shall discuss in Section \ref{subsec:weights}.

\section{Realistic CMB noise and foregrounds}
\label{sec:ILC}
The RS effect appears to be too small to be detected by current and near-term future CMB experiments such as the Simons Observatory (SO) \cite{2019JCAP...02..056A}. Here we consider two future projects: CMB-S4 and a more ambitious and futuristic CMB-HD. The low noise and high resolution of these experiments make them excellent candidates for this kind of analysis. 

CMB-S4 \cite{CMBS4} is a next-generation Stage IV ground based CMB experiment which will reach arcminute resolution and $\approx 1\mu$K-arcmin sensitivity over more than $50\%$ of the sky. The sensitivity and resolution for each map\footnote{From the CMB-S4 wiki \url{https://cmb-s4.uchicago.edu/wiki/index.php/Survey_Performance_Expectations}} for the Large Area Survey is summarized in Table \ref{tab:S4}.

CMB-HD \cite{Sehgal:2019ewc} is a proposed future super-high resolution CMB experiment that utilizes a $\sim 30$m-class dish to achieve $\approx 15$ arcsec resolution and $\approx 0.5 \mu$K-arcmin noise at 150 GHz.  The assumed specifications are listed in Table \ref{tab:HD}, over 50\% of the sky.

\begin{table}[h!]
\begin{tabular}{|c|c|c|c|c|c|c|c|}
\hline
Frequency (GHz)    & 20   & 27   & 39  & 93  & 145 & 225 & 280  \\ \hline
beam FWHM (arcmin) & 10.0   & 7.4  & 5.1 & 2.2 & 1.4 & 1.0 & 0.9  \\ \hline
TT noise level ($\mu$K-arcmin)     & 45.9 & 15.5 & 8.7 & 1.5 & 1.5 & 4.8 & 11.5 \\ \hline
\end{tabular}
\caption{Experimental assumptions for CMB-S4 }
\label{tab:S4}
\end{table}

\begin{table}[h!]
\begin{tabular}{|c|c|c|c|c|c|c|}
\hline
Frequency (GHz)    & 30   & 40   & 90  & 150  & 220 & 280  \\ \hline
beam FWHM (arcmin) & 1.3 & 0.94 & 0.41 & 0.25 & 0.17 & 0.13  \\ \hline
TT noise level ($\mu$K-arcmin)     & 6.5 & 3.4 & 0.73 & 0.79 & 2.0 & 4.6 \\ \hline
\end{tabular}
\caption{Experimental assumptions for CMB-HD}
\label{tab:HD}
\end{table}

Since the signal is extracted by cross-correlating a CMB map with LSS tracers and that the RS effect, like all other gravitational secondary anisotropies, is independent of frequency, we can maximize the $S/N$ of the detection, by minimizing the total noise (or variance) in the CMB map, including the effect of both detector noise and foregrounds. This minimum-variance multi-frequency combination is the standard Internal Linear Combination (ILC, \cite{Eriksen_2004, Tegmark_2003, WMAP:2003cmr}). We use the ILC implementation in the \texttt{BasicILC} code\footnote{\url{https://github.com/EmmanuelSchaan/BasicILC}}. We include the power spectrum of foregrounds in the calculation, using the sky model from \cite{2013JCAP...07..025D}, which includes templates for the angular power spectrum of all of the foregrounds components considered here, as well as their amplitude.
We include all of the relevant small-scale contributions, namely from the thermal and kinematic Sunyaev-Zel'dovich effects (tSZ and kSZ), and well as from the Cosmic Infrared Background (CIB), and radio point sources. Since we are only interested in small scales, we will neglect any effects from the Earth's atmosphere on the noise budget. We also note that the foreground noise can be reduced by masking individually-detected sources beyond the thresholds of \cite{2013JCAP...07..025D}, which should be possible given the better sensitivity of CMB-S4 or CMB-HD compared to the assumptions in that paper: in this sense, our forecasts are conservative because we don't assume any extra masking beyond what has already been demonstrated. It might potentially even make the signal detectable by nearer term experiments such as SO, and it would be worthwhile exploring this further. On the other hand, aggressive masking can also mask part of the signal, potentially lowering the $S/N$ and biasing our interpretation of the signal. We leave a detailed study of the effect of masking on the signal and on the noise to future work.

\begin{figure}
    \centering
    \includegraphics[width=\columnwidth]{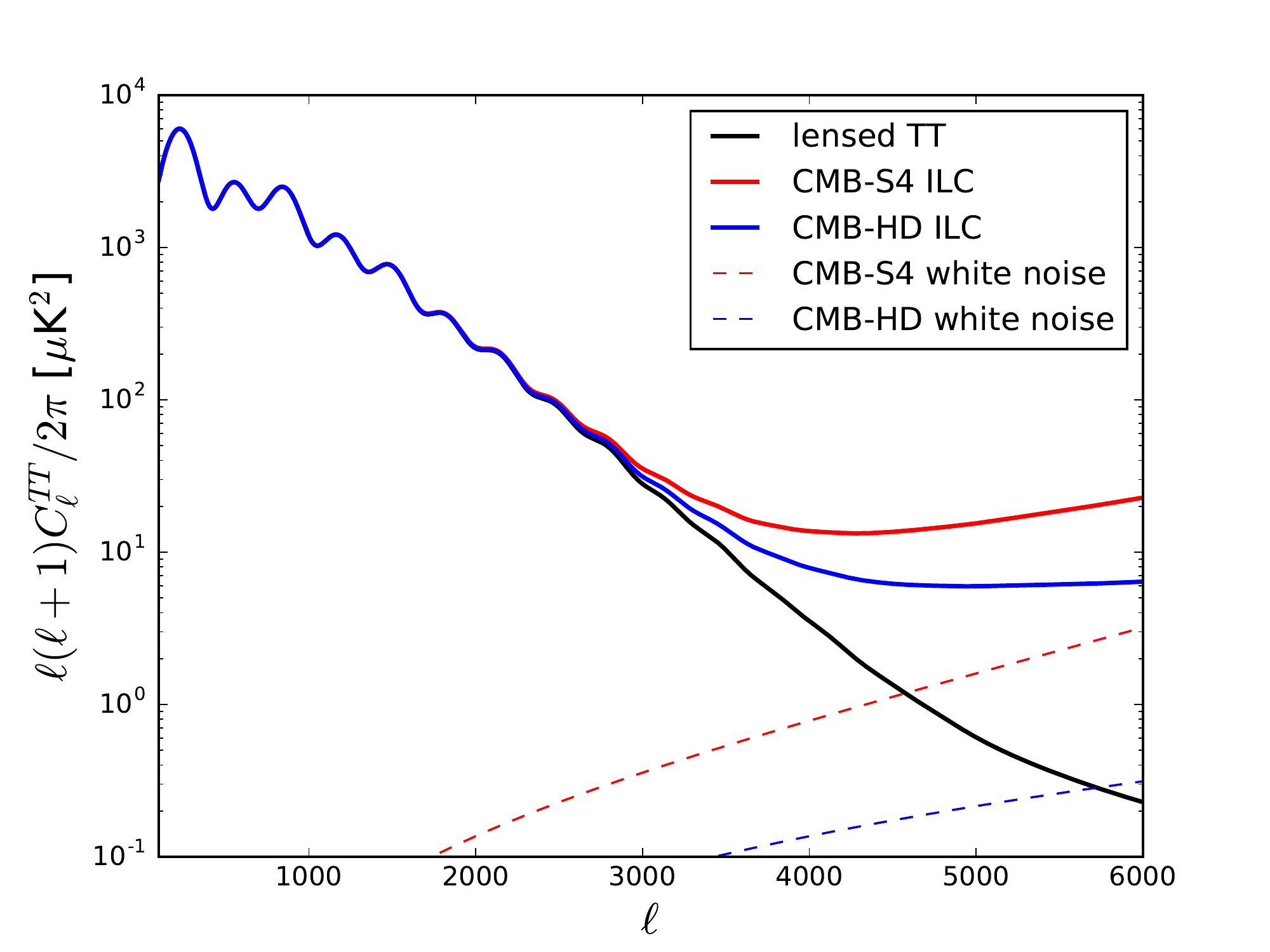}
    \caption{The lensed primary CMB with no foregrounds or noise (black curve), and the total ILC power for CMB-S4 (red) and CMB-HD (blue). For comparison, we plot the white noise only contribution at 145 GHz and 150 GHz for CMB-S4 and CMB-HD respectively, with sensitivities and beams from Tables \ref{tab:S4} and \ref{tab:HD}. We note that the ILC power is considerably larger than even the single frequency noise, due to the additional power from foregrounds.}
    \label{fig:ILC noise}
\end{figure}

From this point on, the white noise of Equation \ref{eq:N_det} will be replaced by the more realistic ILC power as shown in Figure \ref{fig:ILC noise}, which includes the effects of all the known foregrounds.

\section{Large Scale Structure Tracers}
\label{sec:LSS}
In this section we consider a realistic galaxy survey with redshift distribution $n(z)$ and projected number (per steradian) $\bar{n}_{\rm 2D}$. In this general case we have:
\begin{equation}
    \left( \frac{S}{N} \right)^2(\ell_{\rm max}) \approx f_{\rm sky} \sum_{\ell_{\rm min} = 500}^{\ell_{\rm max}} \frac{(2\ell+1) (C_\ell^{Tg})^2} {C_\ell^{gg}C_\ell^{TT}+(C_\ell^{Tg})^2}
    \label{eq:SNR_general}
\end{equation}
where the galaxy power spectrum is given by:
\begin{equation}
    C_{\ell}^{gg} = \int dz \frac{H(z)}{\chi^2} \left [W^{g}(z)\right]^2 P^{\rm NL}_{\delta}(k = \ell/\chi, z) + \frac{1}{\bar{n}_{\rm 2D}}
    \label{eq:gg_general}
\end{equation}
includes shot noise $1/\bar{n}_{\rm 2D}$ in addition to the clustering term.

\subsection{Galaxy Sample}
\label{subsec:galaxies}
\begin{figure}[h!]
\includegraphics[width=\columnwidth]{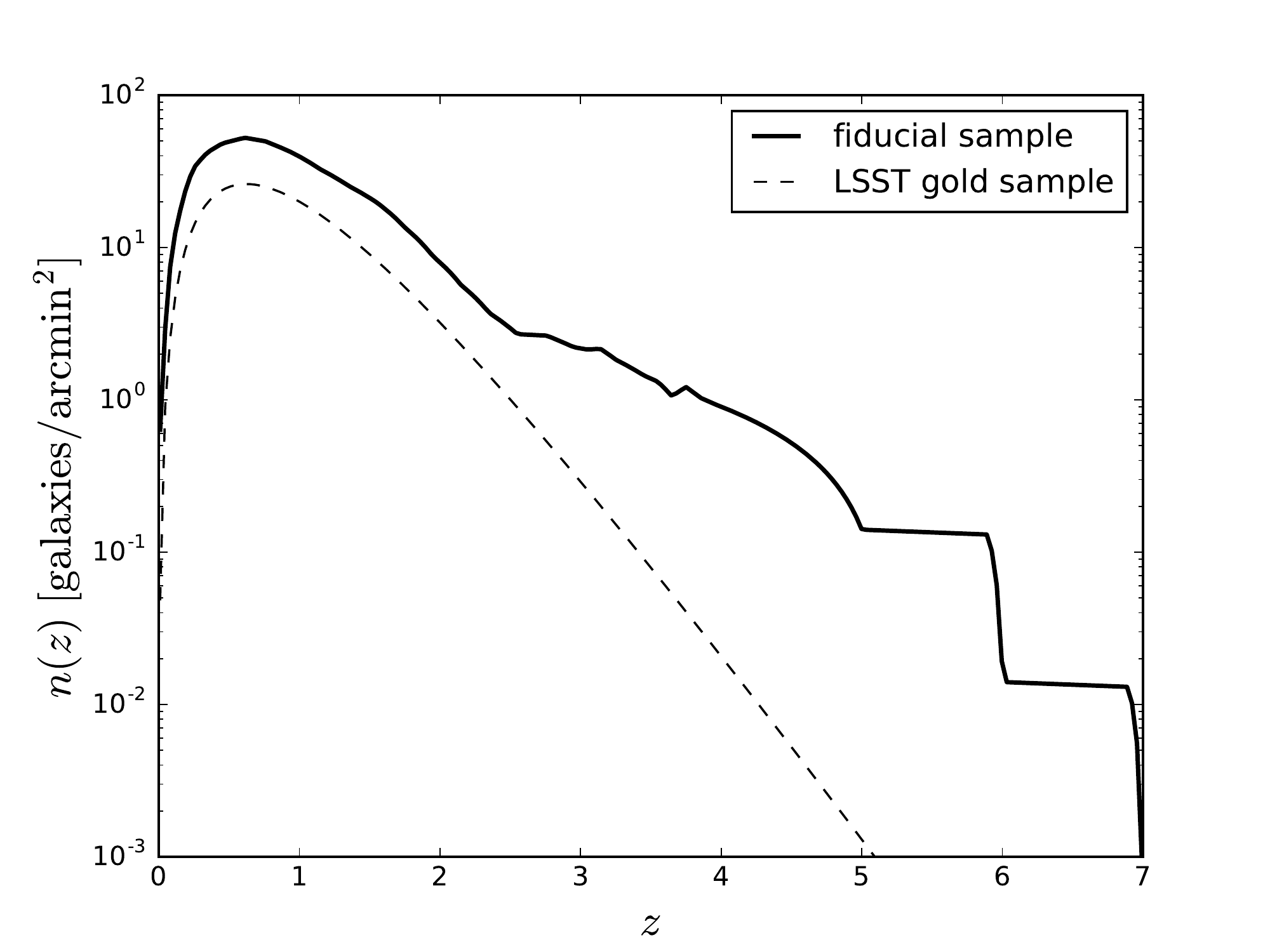}
\caption{Redshift distribution $n(z)$, showing the comparison between the LSST ``gold'' sample with $i < 25.3$ and $\bar{n}_{\rm 2D} \approx 30$ galaxies/arcmin$^2$ and our fiducial sample which includes dropouts, as described in the text.}
\label{fig:dNdz}
\end{figure}

Given the importance of high redshift tracers, we consider a deep sample from LSST, chosen to maximize the number of high-redshift galaxies.
As our fiducial sample we follow the sample defined in \cite{Schmittfull:2017ffw, Yu:2018tem, Yu:2021vce} based on a $i < 27$ magnitude cut with $S/N > 5$.
We also add Lyman break galaxies from
redshift dropouts, whose number density we estimate by scaling recent HSC observations \cite{Ono:2017wjz, Harikane:2017lcw}, yielding $\bar{n}_{\rm 2D} \approx 66$ galaxies/arcmin$^2$ in the range  $z = 0-7$. The inclusion of dropout samples increases the number density by over an order of magnitude at $z \gtrsim 4$ and will have a large impact on the detectability of the RS effect. In Figure \ref{fig:dNdz} we show $n(z)$ for our fiducial sample. We note that such a deep sample may not be suitable for shape measurements for galaxy lensing. However, the angular positions can still be used for measuring clustering and cross-correlation. Similarly, the redshifts may be more poorly constrained (especially for the dropout samples), but we note that the RS kernel is very broad in redshift, and therefore redshift uncertainties are unlikely to affect our conclusions. Following \cite{LSSTSRD, LSSTSciBook}, we take $b_g = 0.95 / G(z)$, where $G(z)$ is the linear growth factor normalized such that $G(z) = 1/(1+z)$ in matter domination.

\subsection{Optimal redshift weights}
\label{subsec:weights}
We note from Figure \ref{fig:dNdz} that $n(z)$ for our fiducial sample peaks at $z < 1$, even with our choices of including deep and dropout samples to maximize the number of high redshift galaxies. Since the ISW kernel peaks at $z \sim 3$, this $n(z)$ is not well-matched to the detection of RS. We solve this problem by deriving optimal redshift weights to apply to each galaxy before cross-correlating with the CMB. 
We find that this weighting is \textit{crucial} to the detectability of the signal.

In detail, before forming the galaxy overdensity map $\delta_g$, we weigh each galaxy by $w(z)$ based on its photometric redshift. As shown in Appendix \ref{app:weights}, the optimal weight around multipole $\ell_*$ is given by
\begin{equation}
    w(z) \propto  \frac{b_g(z) W^{\rm ISW} (k = \ell_*/\chi, z) P^{\rm NL}_{\delta}(k = \ell_*/\chi, z)}{(n(z) / \bar{n}_{\rm 2D}) \left[b^2(z) P^{\rm NL}_{\delta}(k = \ell_*/\chi, z) + 1/n_{\rm 3D}(z) \right]}
    \label{eq:weights}
\end{equation}
where we have defined the 3D number density 
\begin{equation}
    n_{\rm 3D}(z) \equiv \frac{H(z)}{\chi^2(z)} n(z)
\end{equation}
The optimal weights in principle depend on the value of $\ell_*$ chosen above. We choose a fixed $\ell_* = 5000$, since that's where most of the signal comes from, and have confirmed that different choices of $\ell_*$ make virtually no difference to our forecasts. 
For simplicity, we'll leave the weights unnormalized, since they appear with the same power in both the numerator and denominator of the $S/N$ calculation. Hence an overall (redshift-independent) normalization will cancel out. 

The weights have an intuitive explanation: as discussed in Section \ref{sec:where_signal}, most of the signal arises from the range $z \sim 2-4$, with very little contributions from $z \lesssim 1$. However, most galaxy surveys, including Rubin LSST, have the majority of the galaxies at those low redshift, which contribute very little to the signal, but introduce unnecessary noise. Therefore, optimal weights will tend to up-weight galaxies at higher redshift, but this comes at the cost of increasing the effective shot noise (because higher-redshift galaxies are more rare). Optimal weights balance these two competing effects, to maximize the total $S/N$. 

Note that alternatively we could also have divided the sample in several tomographic bins, computed the covariance between them and employed a likelihood function for combining them optimally. The results would be equivalent, but we find our procedure with redshift weights to be more transparent and easier to implement.

With optimal weighting, the calculation of $S/N$ in Equations \ref{eq:SNR_general} and \ref{eq:gg_general} is modified in the following way:
First, the galaxy kernel becomes $W^g(z) \rightarrow w(z) W^g(z)$ in the calculation of $C_{\ell}^{Tg}$ and $C_{\ell}^{gg}$. Moreover, the effective shot noise changes from the original $1/\bar{n}_{\rm 2D}$ to 
\begin{equation}
    \frac{1}{\bar{n}_{\rm 2D}^{\rm eff}} = \frac{1}{\bar{n}_{\rm 2D}^2} \int dz\ n(z) w^2(z)
\end{equation}
As explained above, since the weights are left unnormalized here, the $C_{\ell}^{Tg}$ and $C_{\ell}^{gg}$ with this weighting are only defined up to an arbitrary multiplicative constant, which will cancel in the $S/N$ calculation and hence won't affect any of our results.

\begin{figure}[H]
\includegraphics[width=\columnwidth]{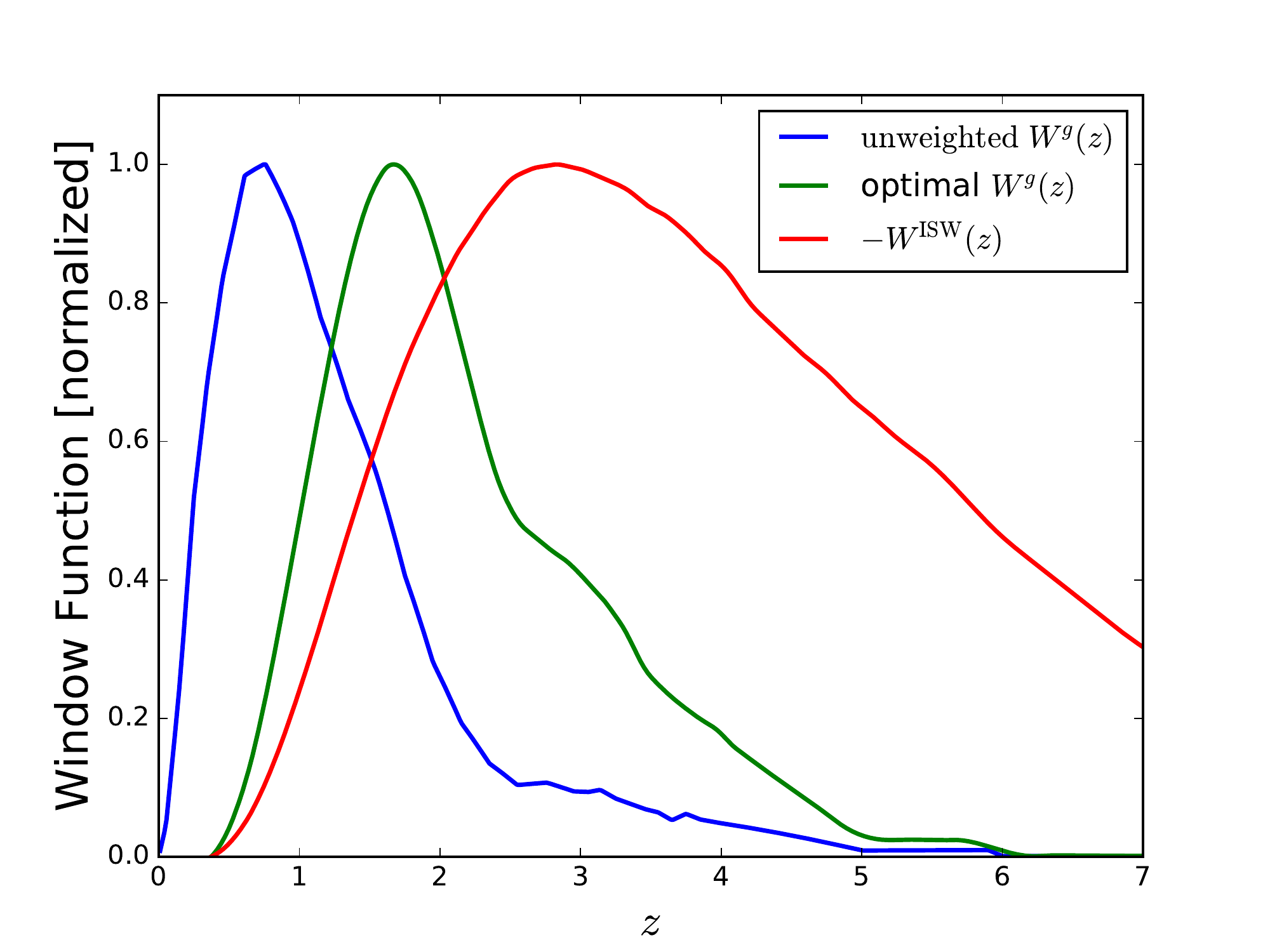}
\caption{Original (unweighted) $W^g(z)$, compared to the optimally weighted $W^g(z)$ for our fiducial sample from LSST, and the ISW kernel, which would represent the optimal weighting for a LSS experiment with zero shot noise. $W^{\rm ISW}(z)$ and the weights evaluated at a fixed multipole $\ell_* = 5000$.}
\label{fig:weights}
\end{figure}
Figure \ref{fig:weights} shows the unweighted (original) galaxy kernel $W^g(z) \propto b_g(z) n(z)$, together the ISW kernel and the optimally weighted kernel $W^g_{\rm opt}(z) \propto w(z) b_g(z) n(z) $. As expected, the reweighted distribution is intermediate between the original $W^g(z)$ at lower redshift (which minimizes the shot noise), and the ISW kernel (which maximizes the signal).

We also note that the optimal weights are smooth in redshift, and therefore we expect photometric redshifts from future imaging surveys to be more than adequate for our purposes. A large fraction of catastrophic redshift outliers can make the weighting scheme suboptimal and bias the interpretation of the signal. Spectroscopic follow-up to LSST, targeting high redshift galaxies, have been proposed \cite{Ferraro:2019uce, Ferraro:2022cmj, Schlegel:2019eqc, Ellis:2019gnt, Marshall:2019wsa} and could provide important calibration, mitigating concerns about photometric redshifts.

\section{Results}
\label{sec:results}
Here we use the CMB noise curves including foregrounds and multi-channel ILC from Section \ref{sec:ILC} and our fiducial galaxy sample described in Section \ref{subsec:galaxies}, together with the optimal weights discussed in Section \ref{subsec:weights}, to forecast the detectability of the signal in a realistic situation. 
\begin{figure}[H]
\includegraphics[width=\columnwidth]{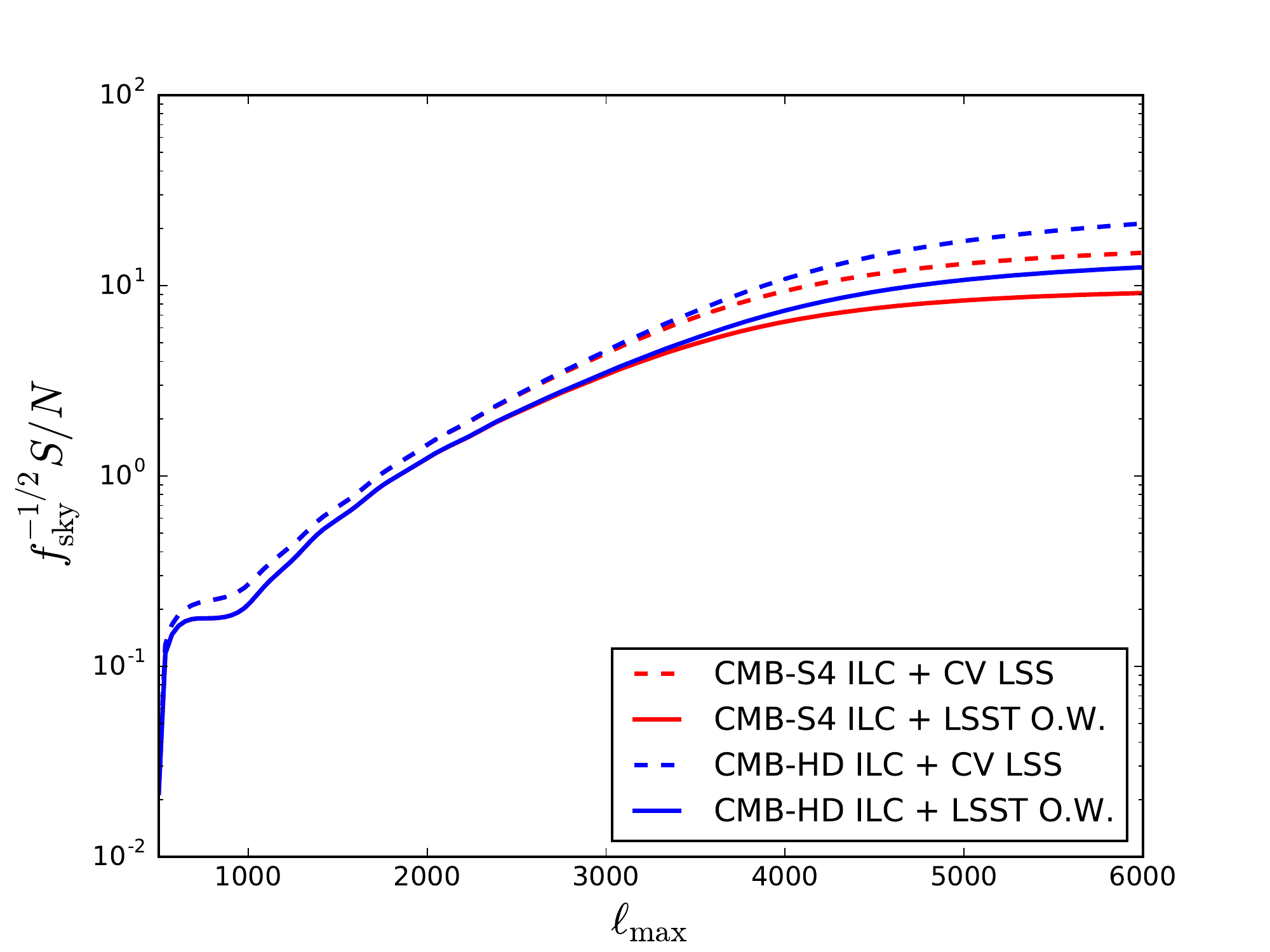}
\caption{The solid lines represent the $S/N$ of the RS signal for CMB-S4 (red) and CMB-HD (blue) in cross-correlation with our fiducial galaxy sample with optimal weighting (O.W.). The dashed lines show what can be obtained with a cosmic variance (CV) limited LSS experiments, in cross correlation with CMB-S4 or CMB-HD. Note that the fiducial surveys are expected to have an overlap over $f_{\rm sky} \approx 0.4$, assuming the current survey strategy.}
\label{fig:SNR_optimal}
\end{figure}
Figure \ref{fig:SNR_optimal} shows the the combination of CMB and our optimally-weighted fiducial sample should lead to a detection of the RS effect with $S/N$ of a few. Assuming an overlap area with $f_{\rm sky} = 0.4$, we find $S/N \approx 6$ for CMB-S4. The more futuristic CMB-HD could improve the detection by $\approx 30\%$ and potentially more by going to higher $\ell_{\rm max}$.

We also note that the optimally weighted fiducial sample is rather close to cosmic variance (within a factor of a few tens of percent), and therefore increasing the density of a LSS survey or changing its redshift distribution is unlikely to lead to large improvements. Significantly increasing the detection significance would require lowering the CMB effective noise, which includes the effect of foregrounds, or going to higher $\ell_{\rm max}$. This can appear challenging, since the foreground power (after ILC) is a weak function of the CMB detector noise. Nonetheless, more aggressive masking than what assumed here can lower the foreground power and should be able to increase the $S/N$, potentially significantly: therefore quantifying this is a worthwhile investigation for future work. 



\section{Conclusions}
\label{sec:conclusions}
We have investigated the detectability of the RS effect with the next generation of high-resolution CMB experiments, when combined with Rubin LSST galaxies. We have shown that in absence of CMB foregrounds, the RS effect is potentially detectable at very high significance; however, when realistic foregrounds are included, the detection significance is limited to $S/N \sim $ 6 - 8, at least in our conservative setting with $\ell_{\rm max} = 6000$. Improving the galaxy sample can only lead to a modest increase in $S/N$ because our galaxy sample is already close to the cosmic variance limit, while significant improvements are possible if the total power of the CMB maps, including foregrounds, can be reduced (for example by more aggressive masking). 

One issue not explored in this paper is the potential bias from foregrounds to the cross-correlation $C_\ell^{Tg}$ itself: foregrounds are always included in the noise, but any residual foreground after ILC might introduce a bias to the inferred RS signal. Studying this quantitatively would require correlated LSS and CMB simulations with high-density samples to high ($z \gtrsim 4$) that are not available to us, and therefore we leave this study to future work. In a real analysis, several tests are possible to ensure that this potential bias is not a concern. For example, one could check that the correct amplitude of the linear ISW is measured on large scales, and that the signal is null in the intermediate regime at $\ell$ of a few hundred. Moreover, imperfect foreground control due to CIB or tSZ contamination is expected to have opposite signs in this cross-correlation: nulling either in turn and checking for consistency would add robustness to the analysis.
We leave this important analysis of foreground biases to future work.

\section*{Acknowledgments}
We thank Matthew Johnson and Mat Madhavacheril for useful discussions. S.F. is supported by the Physics Division of Lawrence Berkeley National Laboratory.
E.S. is supported by the Chamberlain fellowship at Lawrence Berkeley National Laboratory. E.P. is supported by NSF grant AST-1910678 and NASA award 80NSSC18K0403. E.P. is also a Simons Foundation Fellow.
This work used resources of the National Energy Research Scientific Computing Center, a DOE Office of Science User Facility supported by the Office of Science of the U.S. Department of Energy under Contract No. DE-AC02-05CH11231. This work was performed in part at Aspen Center for Physics, which is supported by National Science Foundation grant PHY-1607611.

\bibliographystyle{h-physrev.bst}
\bibliography{refs}


\appendix

\section{Derivation of optimal redshift weights}
\label{app:weights}

Here we derive the optimal weights $w(z)$, introduced in Section \ref{subsec:weights}, and in particular in Equation \ref{eq:weights}. For each multipole $\ell$, we want to maximize the $S/N$ given by each term in the sum in Equation \ref{eq:SNR_general}, where the ``signal'' is given by $S = C_\ell^{Tg}$ and the ``noise'' by $N^2 = C_\ell^{gg} C_\ell^{TT} + (C_\ell^{Tg})^2 \approx C_\ell^{gg} C_\ell^{TT}$, since $T$ and $\delta_g$ are weakly correlated, and $C_\ell^{gg}$ includes the contribution from shot noise. Since most of the $S/N$ comes from $\ell$ of a few thousand, we shall derive the weights around a fixed multipole $\ell_* = 5000$, and we have checked numerically that our results are stable under different choices of $\ell_*$. 

The redshift weights $w(z)$ that we are seeking are to be applied to every galaxy before forming the galaxy overdensity map used in the analysis. When accounting for the weights, Equations \ref{eq:RS_Limber} and \ref{eq:gg_general} for $C_\ell^{Tg}$ and $C_\ell^{gg}$, can be rewritten as:
\begin{align}
    C_{\ell_*}^{Tg} &= \int dz \ g(z) w(z) \\
    C_{\ell_*}^{gg} &= \int dz \ f(z) w^2(z) \label{eq:app_gg}
\end{align}
where
\begin{equation}
   g(z) = \frac{H(z)}{\chi^2} W^{g}(z) W^{\rm ISW}(k = \ell_*/\chi, z) P^{\rm NL}_{\delta}(k = \ell_*/\chi, z)
\end{equation}
and 
\begin{equation}
    f(z) = \frac{H(z)}{\chi^2} \left [W^{g}(z)\right]^2 P^{\rm NL}_{\delta}(k = \ell_*/\chi, z) + \frac{n(z)}{\bar{n}_{\rm 2D}^2}
    \label{eq:app_f}
\end{equation}

We can simplify the calculation by noting that maximizing the $S/N$ is equivalent to minimizing the noise (or its square, since it's a positive quantity), subject to keeping the signal fixed to a fiducial value $S_{\rm fid}$. This constrained minimization problem can be rewritten as an unconstrained minimization with the use of a Lagrange multiplier $\lambda$, and minimizing a ``Lagrangian'' function $\mathcal{L}$ 
\begin{equation}
    \mathcal{L} = N^2 - \lambda [S - S_{\rm fid}]
\end{equation}
with respect to the weights $w(z)$.
Asking for the Lagrangian to be stationary with respect to the weights at a given redshift $z_0$, we get
\begin{equation}
    \frac{\delta \mathcal{L}}{\delta w(z_0)} = 2C_{\ell_*}^{TT} f(z_0) w(z_0) - \lambda g(z_0)= 0 
\end{equation}
from which we get 
\begin{equation}
    w(z_0) = \frac{\lambda}{2C_{\ell_*}^{TT}} \frac{g(z_0)}{f(z_0)} \propto \frac{g(z_0)}{f(z_0)}
\end{equation}
Since the prefactor $\lambda/(2C_{\ell_*}^{TT})$ does not depend on redshift, we are going to drop the normalization and use arbitrarily normalized weights. The normalization drops out in the calculation of $S/N$ and therefore doesn't affect any of our results. Using the definitions of $f$ and $g$ above and noting that $W^g(z) = b_g(z) n(z) /\bar{n}_{\rm 2D}$, we get
\begin{equation}
     w(z_0) \propto  \frac{b_g(z_0) W^{\rm ISW} (k = \ell_*/\chi_0, z_0) P^{\rm NL}_{\delta}(k = \ell_*/\chi_0, z_0)}{(n(z) / \bar{n}_{\rm 2D}) \left[b^2(z_0) P^{\rm NL}_{\delta}(k = \ell_*/\chi_0, z_0) + 1/n_{\rm 3D}(z_0) \right]}
     \label{eq:app_weights}
\end{equation}
where $\chi_0 = \chi(z_0)$ and we have defined the 3D galaxy density in terms of the 2D redshift distribution as:
\begin{equation}
    n_{\rm 3D}(z) \equiv \frac{H(z)}{\chi^2(z)} n(z)
\end{equation} 
Equation \ref{eq:app_weights} is precisely Equation \ref{eq:weights} in the main text. Moreover, the comparison between Equation \ref{eq:app_gg} and \ref{eq:app_f} makes it clear that the shot noise for the weighted field is modified in the following way:
\begin{equation}
    \frac{1}{\bar{n}_{\rm 2D}^{\rm eff}} = \frac{1}{\bar{n}_{\rm 2D}^2} \int dz\ n(z) w^2(z)
\end{equation}
and hence this concludes the derivation.
\end{document}